\documentclass[aps,pra,reprint,showpacs,superscriptaddress,twocolumn,notitlepage,10pt]{revtex4-1}%
\usepackage{ulem}
\usepackage{amsfonts}
\usepackage{mathrsfs}
\usepackage{amsmath}
\usepackage{amssymb}
\usepackage{graphicx}
\usepackage{dcolumn}
\usepackage{color}%
\usepackage[colorlinks,linkcolor=blue,citecolor=blue,hyperindex,bookmarks=false,pdfstartview=FitH]{hyperref}
\newcommand{\red}[1]{\textcolor[rgb]{0.80,0.00,0.00}{#1}}

\newcommand{\figpanel}[2]{\hyperref[#1]{\ref*{#1}(#2)}}

\begin{document}
\title{Quantum transport in nonlinear Rudner-Levitov models}
\author{Lei Du}
\affiliation{Beijing Computational Science Research Center, Beijing 100193, China}
\author{Jin-Hui Wu}
\affiliation{Center for Quantum Sciences and School of Physics, Northeast Normal University, Changchun 130024, China}
\author{M. Artoni}
\email{artoni@lens.unifi.it}
\affiliation{Department of Engineering and Information Technology and INO-CNR Sensor Lab, Brescia University, 25133 Brescia, Italy}
\author{G. C. La Rocca}
\email{larocca@sns.it}
\affiliation{NEST, Scuola Normale Superiore, 56126 Pisa, Italy}
\date{\today }

\begin{abstract}
Quantum transport in a class of nonlinear extensions of the Rudner-Levitov model is numerically studied in this paper. We show that the quantization of the mean displacement, which embodies the quantum coherence and the topological characteristics of the model, is markedly modified by nonlinearities. Peculiar effects such as a ``trivial-nontrivial'' transition and unidirectional long-range quantum transport are observed. These phenomena can be understood on the basis of the dynamic behavior of the effective hopping terms, which are time and position dependent, containing contributions of both the linear and nonlinear couplings.d nonlinear couplings. \end{abstract}
\maketitle

\section{Introduction}

The Rudner-Levitov (RL) model~\cite{RLmodel}, which can be viewed as a simple non-Hermitian extension of the Su-Schrieffer-Heeger (SSH) model~\cite{SSH}, has been predicted to support quantized quantum transport with topological protection~\cite{RLmodel,RLarxiv} has been verified experimentally in a lattice of optical waveguides~\cite{RL2015}. This important feature not only shows the possibility of accessing nontrivial topological phases in physical systems where dissipations are intrinsic and unavoidable, but also provides a direct way to detect the topology of such systems by monitoring their bulk properties. Specifically, for a dimerized ladder of discrete quantum states, with one sublattice lossy and the other one neutral, the mean displacement of a particle which is initialized at a neutral site shows well quantized behaviors, depending on the ratio between the intra- and inter-cell coupling constants. 

Recently, the intersection of nonlinear optics and topological photonics has unlocked a series of peculiar effects that have no counterparts in condensed matter topologies~\cite{NonTopoRev}, such as magnet-free nonreciprocity~\cite{MFnr1,MFnr2}, active tunability~\cite{tuna1,tuna2,tuna3}, self-induced topological transitions~\cite{Hadad,Hadad2,Ezawa}, topological solitons~\cite{tsoliton1,tsoliton2,tsoliton3,tsoliton4,chong2021}, and topological frequency conversions~\cite{FC1,FC2,FC3}. Most recently, Xia \textit{et al.}  have considered a nonlinear PT-symmetric optical waveguide lattice to investigate the mutual interplay between topology, PT symmetry, and nonlinearity~\cite{XiaScience}, demonstrating that global effects, i.e., topological and PT transitions, can be actively controlled by the local nonlinearity. The RL model has already been generalized by including on-site Kerr-type nonlinearities, which shows that the quantized quantum transport is severely affected by the nonlinearity induced decoherence effects~\cite{Rapedius} and that the nonlinear effect becomes pronounced in the PT-broken region~\cite{RLSR}. However, the study of the influence of nonlinearity on the quantum transport in the RL model is still at the initial stage. In particular, work on how nonlinearities affect the hopping terms, rather than the on-site energies, is still missing; the latter in fact may affect quantum transport in the RL model precisely because its topological properties are dictated by the ratio between the intra- and inter-cell coupling constants.

In this paper, we consider specific extensions of the RL model with different types of nonlinear couplings and numerically study the quantum transport therein. As it happens for the case of on-site nonlinearities~\cite{Rapedius,RLSR}, our results show that the nonlinear coupling generates departures from the typical quantized behavior of the particle transport. Our hopping nonlinearities turn out to be responsible for new effects, such as (\textit{i.}) a ``trivial-nontrivial'' transition, by which the mean displacement will always take the value of $0$ or $1$ regardless of the linear coupling imbalance, and (\textit{ii.}) a long-range transport, by which the particle can move unidirectionally across many unit cells before its decay. These phenomena can be understood on the basis of the dynamic behavior of the effective time and position dependent hopping terms containing contributions of both the linear and nonlinear couplings. Our findings are relevant to experimental platforms, such as arrays of coupled optical waveguides and cavities, but may also be adapted to acoustic or electric circuit resonators as well as superconducting quantum circuits~\cite{Kono,Smirn,Xue}.

\section{The Rudner-Levitov model}

\begin{figure}[ptb]
\includegraphics[width=8.5cm]{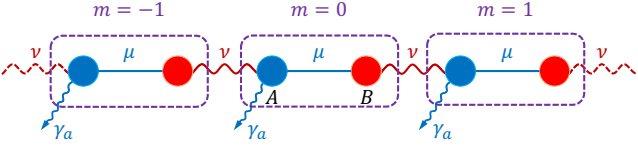} 
\caption{(Color online) Schematic illustration of the Ruder-Levitov model, where $\mu$ ($\nu$) and $\gamma_{a}$ are the intracell (intercell) coupling constant and the decay rate of the lossy sites, respectively.}
\label{fig1}
\end{figure}

We start by briefly reviewing the original RL model~\cite{RLmodel,RLarxiv,RL2015}, where each unit cell of the one-dimensional dimerized lattice contains a lossy site $A$ and a neutral site $B$, as shown in Fig.~\ref{fig1}. The intra- and inter-cell coupling constants are denoted by $\mu$ and $\nu$, respectively, while the decay rate of each lossy site is denoted by $\gamma_{a}$. In the lossless limit $\gamma_{a}\rightarrow0$, such a linear bipartite lattice is equivalent to the SSH model~\cite{SSH} which shows a topological transition depending on the ratio $\mu/\nu$~\cite{zak,asboth}. It has been shown that the nonvanishing loss in the RL model plays the key role for observing quantized quantum transport. The real-space Hamiltonian of the RL model is given by ($\hbar=1$ hereafter)
\begin{equation}
\begin{split}
H_{0}&=\sum_{m}\Big[-i\gamma_{a}|m,a\rangle\langle m,a|-(\nu |m,a\rangle\langle m-1,b|\\
&\quad\,-\mu |m,a\rangle\langle m,b|+\text{H.c.})\Big],
\end{split}
\label{eq1}
\end{equation}
where $m\in[-N,N]$ labels the unit cells of the lattice; $|m,a\rangle$ and $|m,b\rangle$ denote the Wannier states of sites $A$ and $B$ in the $m$th unit cell, respectively.

Setting $|\psi(t)\rangle=\sum_{m}[a_{m}(t)|m,a\rangle+b_{m}(t)|m,b\rangle]$ as a general eigenstate of the system, with $a_{m}$ ($b_{m}$) the field amplitude of $A$ ($B$) in the $m$th unit cell, it is straightforward to attain the following dynamic equations
\begin{equation}
\begin{split}
\frac{\partial a_{m}}{\partial t}  &  =-\gamma_{a}a_{m}+i\nu b_{m-1}+i\mu b_{m},\\
\frac{\partial b_{m}}{\partial t}  &  =i\mu a_{m}+i\nu a_{m+1}.
\end{split}
\label{eq2}
\end{equation}
Hereafter in this paper, we assume $\mu=0.5-\delta g$ and $\nu=0.5+\delta g$ with $\delta g$ denoting the coupling imbalance ($|\delta g|\le 0.5$). Then, one can numerically calculate the mean displacement of a particle before its decay, i.e.,
\begin{equation}
\langle\Delta m\rangle=\sum_{m}m\int_{0}^{\infty}2\gamma_{a}\vert a_{m}(t)\vert^{2}dt.
\label{eq3}
\end{equation}
This is equivalent to the winding number of the relative phase between components of the Bloch wave function [i.e., the eigenstate of the Fourier transformation of $H(m)$] and is thus topologically protected~\cite{RLmodel,RLarxiv,zak,asboth}. Specifically, if one initializes a single particle at a neutral site, the mean displacement $\langle\Delta m\rangle$ takes the value of $0$ for $\nu<\mu$ and takes the value of $1$ for $\nu>\mu$, which precisely predicts the topological phase transition of this model. 


Such an archetype model has been well investigated both theoretically~\cite{RLmodel,RLarxiv} and experimentally~\cite{RL2015}, and has also been extended in many directions~\cite{Rapedius,RLSR,LiExtend,LieuExtend,LonghiExtend,JinExtend,YuceExtend,RLdl,SAP,WuExtend,BergExtend}. In the following, we generalize the RL model to a few instances with nonlinear hopping terms and study the quantum transport therein. 

\section{Nonlinearity induced trivial-nontrivial transition}

We anticipate that all our nonlinear extensions have symmetric nonlinear couplings so that the norm of a quantum state still evolves according to $d\langle\psi|\psi\rangle/dt=-\sum_{m}2\gamma_{a}|a_{m}|^{2}$ as in the linear RL model (see Appendix~\ref{appa}); this validates in turn the definition of the mean displacement in Eq.~(\ref{eq3}). In the absence of losses, this assures the conservation of the standard norm of a quantum state; other nonlinear models, such as those related to the Ablowitz-Ladik system~\cite{NDDE, ALpra,renormAL}, do not have this property and are outside the scope of this paper.

\begin{figure}[ptb]
\includegraphics[width=8.5cm]{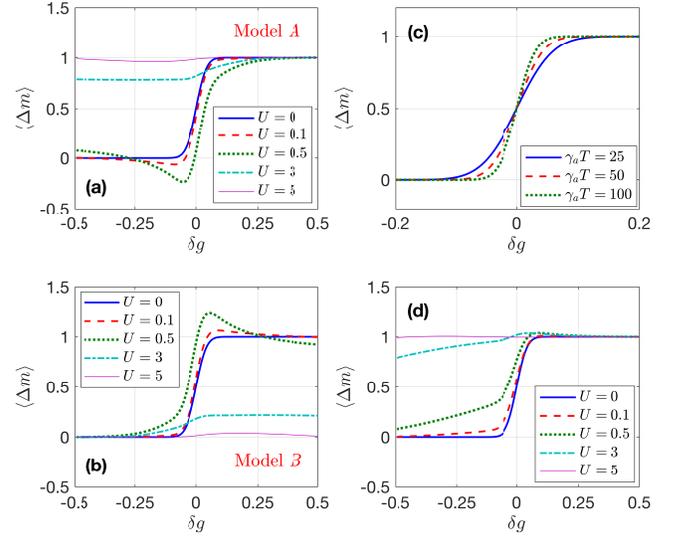} 
\caption{(Color online) Mean displacement $\langle\Delta m\rangle$ of the particle versus coupling imbalance $\delta g$ for (a) model $A$ and (b) model $B$, with initial state $|\psi(t=0)\rangle=|m=0,b\rangle$, integral time $\gamma_{a}T=50$, and different values of nonlinear coefficient $U$. (c) $\langle\Delta m\rangle$ versus $\delta g$ for model $A$ with $|\psi(t=0)\rangle=|m=0,b\rangle$, $U=0$, and different values of integral time $T$. (d) $\langle\Delta m\rangle$ versus $\delta g$ for model $A$, with identical parameters to those in (a) except for $\mu\rightarrow -\mu$ and $\nu\rightarrow -\nu$. Other parameters are $\mu=0.5-\delta g$, $\nu=0.5+\delta g$ and$\gamma_{a}=2$.}
\label{fig2}
\end{figure}
\vspace{2mm}

\textit{Model A.}
First, consider the case in which the \textit{intercell coupling} terms contain a Kerr-type nonlinear part that depends on the field amplitudes of both lossy and neutral sites. The relevant dynamics follows the nonlinear Schr\"{o}dinger equations
\begin{equation}
\begin{split}
\frac{\partial a_{m}}{\partial t}  &  =-\gamma_{a}a_{m}+i\mu b_{m}+i(\nu-\xi_{m})b_{m-1},\\
\frac{\partial b_{m}}{\partial t}  &  =i\mu a_{m}+i(\nu-\xi_{m+1})a_{m+1}.
\end{split}
\label{eq4}
\end{equation}
where $\xi_{m}=U(|a_{m}|^{2}+|b_{m-1}|^{2})$ corresponds to the nonlinear part of the intercell coupling between the $m$th and $(m-1)$th unit cells, the strength of which is described by the nonlinear coefficient $U$. Eq.~(\ref{eq4}) embeds 
a non-Hermitian extension of the model in Ref.~\cite{Hadad} and can be employed to describe arrays of coupled optical and acoustic cavities, as well as circuit resonators.

We plot in Fig.~\figpanel{fig2}{a} the mean displacement $\langle\Delta m\rangle$ as a function of the coupling imbalance $\delta g$ for model $A$ with the initial state $|\psi(t=0)\rangle=|m=0,b\rangle$. In the absence of nonlinearities ($U=0$), the mean displacement of the particle before its decay is quantized with two integer values $0$ and $1$: the particle initially localized at the central neutral site will hop to the lossy site in the right ($m=1$) unit cell and then decays from the lattice if $\nu>\mu$, or it will hop to the left lossy site within the initial ($m=0$) unit cell and then decays if $\nu<\mu$. Note that the transition from $\langle\Delta m\rangle=0$ to $\langle\Delta m\rangle=1$ here is not perfectly quantized due to the finite integration time. As shown in Fig.~\figpanel{fig2}{c}, the quantization behavior becomes more and more ideal with the increase of the integration time $T$. As the nonlinear coefficient increases from zero, the mean displacement deviates from the quantized behavior gradually. In particular, the original ``topological trivial'' region (where $\langle\Delta m\rangle=0$ in the linear case~\cite{RLmodel,RLarxiv}) almost disappears upon increasing $U$, with the values of $\langle\Delta m\rangle$ approaching unity over the whole parametric range. In other words, the particle always hops to the right unit cell for large enough $U$, regardless of the relative values of $\mu$ and $\nu$.

\vspace{2mm}

\textit{Model B.}
We consider next the case in which the \textit{intracell coupling} is instead modified by a Kerr-type nonlinearity. The dynamics can now be described by,
\begin{equation}
\begin{split}
\frac{\partial a_{m}}{\partial t}  &  =-\gamma_{a}a_{m}+i(\mu-\zeta_{m})b_{m}+i\nu b_{m-1},\\
\frac{\partial b_{m}}{\partial t}  &  =i(\mu-\zeta_{m})a_{m}+i\nu a_{m+1},
\end{split}
\label{eq5}
\end{equation}
with $\zeta_{m}=U(|a_{m}|^{2}+|b_{m}|^{2})$ and the corresponding numerical results are plotted in Fig.~\figpanel{fig2}{b}. It can be seen that the mean displacement for model $B$ shows exactly reverse behaviors compared with those for model $A$: upon increasing $U$, the originally ``nontrivial'' region (where $\langle\Delta m\rangle=1$ in the linear case) almost disappears. In this case, the particle always hops from a neutral site to the left lossy one within the same unit cell. 

\begin{figure}[ptb]
\includegraphics[width=8.5cm]{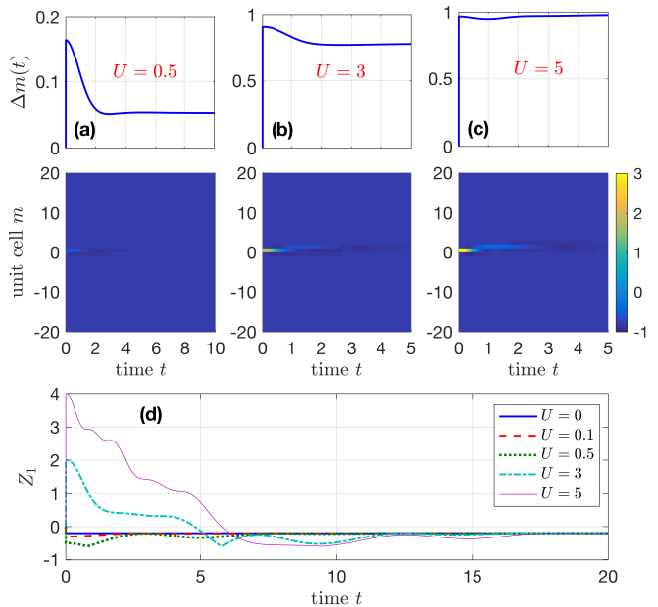} 
\caption{(Color online) Dynamic evolutions of displacement $\Delta m(t)$ (upper) and effective coupling contrast $Z_{m}$ (lower) for model $A$, with (a) $U=0.5$, (b) $U=3$, and (c) $U=5$. (d) Dynamic evolutions of effective coupling contrast $Z_{1}$ with different values of $U$. The initial state is assumed as $|\psi(t=0)\rangle=|m=0,b\rangle$. Other parameters are $\delta g=-0.4$ for (a)-(c), $\delta g=-0.1$ for (d), and $\gamma_{a}=2$.}
\label{fig3}
\end{figure}

The results above can be  understood from the effective inter- and intra-cell couplings which contain the contributions of the nonlinear parts. As an example, we define for model $A$ the effective coupling contrast $Z_{m}=|\nu_{\text{eff},m}|-|\mu|$ with $\nu_{\text{eff},m}=\nu-\xi_{m}$ the effective intercell coupling constant, and plot in Fig.~\ref{fig3} the dynamic evolutions of $Z_{m}$ with different values of $U$. Moreover, we also plot the evolutions of the time-dependent particle displacement $\Delta m(t)=\sum_{m}m\int_{0}^{t}2\gamma_{a}\vert a_{m}(t')\vert^{2}dt'$~\cite{Rapedius} showing how it reaches its final value. In the absence of nonlinearities, the coupling contrast $Z_{m}$ reduces to $Z=|\nu|-|\mu|$, which is position independent and can be used to predict the transition of $\langle\Delta m\rangle$, i.e., $\langle\Delta m\rangle=1$ if $Z>0$ and $\langle\Delta m\rangle=0$ if $Z<0$. In the presence of nonlinearities, however, $Z_{m}$ becomes dependent on $m$. It can be seen from Figs.~\figpanel{fig3}{a}-\figpanel{fig3}{c} that for $\delta g=-0.4$ ($\nu=0.1$), the value of $Z_{m=1}$ increases with $U$ from negative to positive monotonously. On the other hand, for $\delta g=-0.1$ ($\nu=0.4$) as shown in Fig.~\figpanel{fig3}{d}, with the increase of $U$, the value of $Z_{m=1}$ first diminishes until $Z_{m=1}=-|\mu|$, then it turns to increase with $U$ gradually from negative to positive. This is also why the the mean displacement shows a non-monotonic behavior as $U$ increases in this case. In Appendix~\ref{appb}, we provide a detailed explanation for the physical meaning of the positive and negative fractional values of $\langle\Delta m\rangle$. As a consequence, the topology in the ``trivial'' region can be modified by the nonlinearities and thus one can always observe nontrivial quantum transport for large enough $U$. Although the region where the contrast $Z_{m}$ is modified by the nonlinearity disappears with time, its duration is long enough for the particle displacement to reach the final values [see the figures in the upper row of Figs.~\figpanel{fig3}{a}-\figpanel{fig3}{c}]. Intuitively, the influence of the nonlinearities disappears after the particle spreads out and decays from the lattice via the lossy sites. Similarly, the ``nontrivial-to-trivial'' transition of model $B$ can also be understood from the effective coupling contrast. 

In Eqs.~(\ref{eq4}) and (\ref{eq5}), we have assumed that the linear and nonlinear parts of the coupling terms tend to cancel each other, i.e., the signs of $\mu$ and $\nu$ are opposite to that of $U$. Such an assumption is responsible for the non-monotonic behavior of $\langle\Delta m\rangle$ as $U$ changes. For comparison, we reverse the signs of $\mu$ and $\nu$ (i.e., $\mu\rightarrow -\mu$ and $\nu\rightarrow -\nu$) in Eq.~(\ref{eq4}) and plot in Fig.~\figpanel{fig2}{d} $\langle\Delta m\rangle$ versus $\delta g$ in this case. It can be seen that the values of $\langle\Delta m\rangle$ in the originally trivial region increase monotonously with $U$ to approach unity over the whole parametric range. This is different from the non-monotonic behavior in Fig.~\figpanel{fig2}{a}. Such a result is due to the fact that the effective coupling contrast $Z_{m=1}$ increases with $U$ monotonously in this case. 

At the end of this section, we would like to point out that the ``trivial-nontrivial'' transition can also be observed even if the nonlinear coupling terms only depend on the field amplitudes of the neutral sites. To prove this, we consider in Appendix~\ref{appc} such a case and demonstrate the behaviors of the quantum transport therein. 

\section{Nonlinearity induced unidirectional long-range displacement}

\begin{figure}[ptb]
\includegraphics[width=8.5cm]{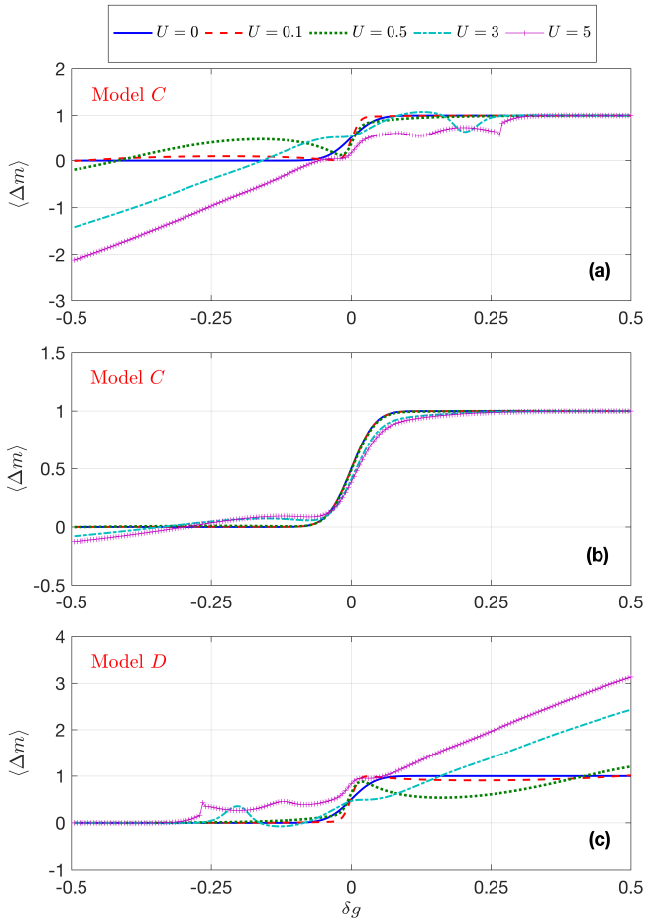} 
\caption{(Color online) Mean displacement $\langle\Delta m\rangle$ of the particle versus coupling imbalance $\delta g$ for [(a) and (b)] model $C$ and (c) model $D$, with initial state $|\psi(t=0)\rangle=|m=0,b\rangle$ and different values of nonlinear coefficient $U$. We assume $\gamma_{a}=0.2$ for (a) and (c), while $\gamma_{a}=2$ for (b).}
\label{fig4}
\end{figure}

\vspace{2mm}

\textit{Model \{C. \& D.\}}
Finally, we consider in this section the case in which the nonlinear coupling constants are only dependent on the field amplitudes of the lossy sites.
The relevant dynamic equations take the form (\textit{Model C.)},
\begin{equation}
\begin{split}
\frac{\partial a_{m}}{\partial t}  &  =-\gamma_{a}a_{m}+i\mu b_{m}+i(\nu-\chi_{m})b_{m-1},\\
\frac{\partial b_{m}}{\partial t}  &  =i\mu a_{m}+i(\nu-\chi_{m+1})a_{m+1}
\end{split}
\label{eq6}
\end{equation}  
with $\chi_{m}=U|a_{m}|^{2}$ describing the nonlinear part of the intercell couplings in this case. Similarly, when the intracell couplings are modified by the field amplitudes of the lossy sites in a nonlinear manner one has for the dynamic equations (\textit{Model $D.$}),
\begin{equation}
\begin{split}
\frac{\partial a_{m}}{\partial t}  &  =-\gamma_{a}a_{m}+i(\mu-\chi_{m})b_{m}+i\nu b_{m-1},\\
\frac{\partial b_{m}}{\partial t}  &  =i(\mu-\chi_{m})a_{m}+i\nu a_{m+1}.
\end{split}
\label{eq7}
\end{equation} 

As in the last section, we first examine the mean displacement $\langle\Delta m\rangle$ of the particle versus the coupling imbalance $\delta g$ for both models $C$ and $D$, as shown in Figs.~\figpanel{fig4}{a}-\figpanel{fig4}{c}. Interestingly, we find that long-range particle displacements, i.e., $\langle\Delta m\rangle<0$ or $\langle\Delta m\rangle>1$, are allowed within these two extensions of the RL model. As shown in Fig.~\figpanel{fig4}{a}, for instance, the value of $\langle\Delta m\rangle$ can be smaller than $-1$ for large $U$, implying that the particle initially located at a neutral site can hop on average several times to the left before its decay if the nonlinearities are strong enough. Note that a relatively small $\gamma_{a}$ ($\gamma_{a}=0.2$) is assumed in this case in order to observe the such unidirectional long-range displacements, while for large $\gamma_{a}$ the mean displacement becomes less sensitive to $U$, as shown in Fig.~\figpanel{fig4}{b}. Moreover, model $D$ displays reverse behaviors of the mean displacement, i.e., the particle will move to the right with $\langle\Delta m\rangle>1$ for large enough $U$. 

\begin{figure}[ptb]
\includegraphics[width=8.5cm]{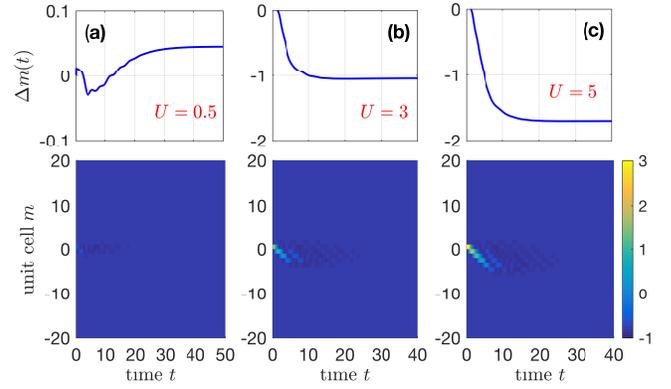} 
\caption{(Color online) Dynamic evolutions of displacement $\Delta m(t)$ (upper) and effective coupling contrast $Z_{m}$ (lower) for model $C$, with (a) $U=0.5$, (b) $U=3$, and (c) $U=5$. The initial state is assumed as $|\psi(t=0)\rangle=|m=0,b\rangle$. Other parameters are $\delta g=-0.4$ and $\gamma_{a}=0.2$.}
\label{fig5}
\end{figure} 

The underlying physics of the above results can be qualitatively understood as follows. For positive $\delta g$ the intercell coupling is stronger than the intracell one, thus the particle initially localized at a neutral site will hop to the right lossy site and eventually decay. In this case, the value of the mean displacement cannot be larger than $1$, because the particle will not keep on moving due to the weak intracell coupling. In contrast, for negative $\delta g$, the particle will hop to the left lossy site within the initial unit cell (i.e., the $0$th unit cell in Fig.~\ref{fig4}). For model $C$, the effective intercell coupling between the $0$th and $-1$th unit cells is enhanced after the particle has moved to the site $|m=0,a\rangle$. If the nonlinearity is strong enough, the particle can keep on moving towards the left and thereby the mean displacement should increase with $U$. Of course, owing to the dissipation at each lossy site, the mean displacement is still finite since the effective intercell coupling will no longer be stronger than the adjacent intracell coupling after the particle has moved across several lossy sites. This is also why the long-range particle displacement tends to vanish if the decay rate $\gamma_{a}$ is large enough. The reverse behaviors in model $D$ can be interpreted in a similar way.  

Similar to the last section, we plot in Figs.~\figpanel{fig5}{a}-\figpanel{fig5}{c} the dynamic evolutions of the time-dependent displacement $\Delta m(t)$ as well as the effective coupling contrast $Z_{m}=|\nu-\chi_{m}|-|\mu|$ of model $C$ with different values of $U$. Once again, one can find that the values of $Z_{m}$ in a certain region increase with $U$. Unlikely in Fig.~\ref{fig3}, however, where only the nearest-neighbor intercell coupling (the coupling between the $0$th and the $1$th unit cells) is markedly modified by the nonlinearities, here model $C$ exhibits a larger modification region that extends toward the left side of the model. This again reflects the origin of the long-range displacements. For all cases in Fig.~\ref{fig5}, the particle displacement reaches its final value before the modification region disappears. 

\begin{figure}[ptb]
\includegraphics[width=8.5cm]{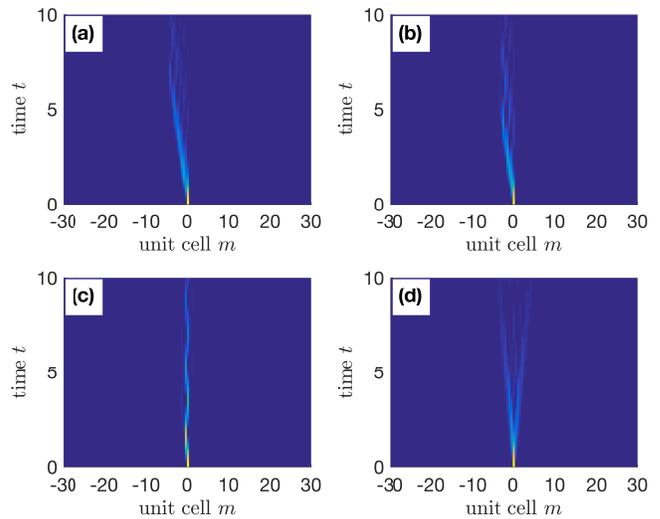} 
\caption{(Color online) Dynamic evolutions of initial state $|\psi(t=0)\rangle=|m=0,b\rangle$ for model $C$ with (a) $U=5$ and $\delta g=-0.4$, (b) $U=3$ and $\delta g=-0.4$, (c) $U=0.5$ and $\delta g=-0.4$, and (d) $U=0.5$ and $\delta g=0$. In all cases, we take $\gamma_{a}=0.2$.}
\label{fig6}
\end{figure} 

The results in Fig.~\ref{fig5} can also be verified by the time evolutions of the initial state $|\psi(t=0)\rangle=|m=0,b\rangle$. As shown in Figs.~\figpanel{fig6}{a}-\figpanel{fig6}{c}, the particle transport tends to be unidirectional as the nonlinearities strengthen. Note that for $\delta g=0$, the lattice is homogeneous at the initial time, such that the particle transport should be almost symmetric, insensitive to the value of $U$. As an example, Fig.~\figpanel{fig6}{d} shows the dynamic evolution of the initial state for $\delta g=0$ and $U=0.5$, which is similar to the symmetric diffusion in a one-dimensional homogeneous lattice. In Fig.~\figpanel{fig6}{c}, however, the excitation is confined within a narrow region around $m=0$ because the effective intercell couplings are extremely weak (both the linear and nonlinear parts are weak) in this case.  

\section{Conclusions}

In summary, we have extended the Rudner-Levitov (RL) model allowing for nonlinear hopping and assessed how nonlinearities affect the mean displacement in quantum transport. At variance with on-site nonlinearities, dynamic modifications of the ratio between the effective intra- and inter-cell couplings directly affect the behavior of quantum transport related to the topological characteristics of the RL model. Our results encompass different extensions of the RL model, which are interpreted qualitatively in terms of the local and time-dependent effective coupling contrast function $Z_{m}(t)$. 
There is one specific extension (See Sect.~IV) whereby for strong nonlinearities and low losses unidirectional long-range quantum transport may emerge. The present work is relevant to experimental platforms that hinge on tight-binding Hamiltonians, including non-Hermitian Hamiltonians and those describing nonlinear effects in optics, in acoustics, and in electronics. 

\section*{Acknowledgments}

This work is supported by the National Natural Science Foundation of China (No. U1930402 and No. 12074061), the Cooperative Program by the Italian Ministry of Foreign Affairs and International Cooperation (MAECI) through Italy-China International Cooperation (PGR00960), and the ``Laboratori Congiunti" 2019 Program of National Research Council (CNR) of Italy.

\appendix
\section{Norm evolution with symmetric couplings}\label{appa}

It has been shown that the norm of a quantum state should evolve according to 
\begin{equation}
\frac{d}{dt}\langle\psi|\psi\rangle=-\sum_{m}2\gamma_{a}|a_{m}|^{2}
\label{A1}
\end{equation}   
to guarantee the validity of the definition of the mean displacement in Eq.~(\ref{eq3})~\cite{RLmodel}. In this appendix, we would like to prove that Eq.~(\ref{A1}) still holds for nonlinear RL models with symmetric couplings. For example, according to Eq.~(\ref{eq4}) in the main text, the norm of a quantum state in model $A$ should follows
\begin{equation}
\begin{split}
\frac{d\langle\psi|\psi\rangle}{dt}&=(\frac{d}{dt}\langle\psi|)|\psi\rangle+\langle\psi|(\frac{d}{dt}|\psi\rangle)\\
&=\sum_{m}\Big[(\frac{da_{m}^{*}}{dt})a_{m}+(\frac{db_{m}^{*}}{dt})b_{m}\\
&\quad\,+a_{m}^{*}(\frac{da_{m}}{dt})+b_{m}^{*}(\frac{db_{m}}{dt})\Big]\\
&=\sum_{m}\Big[-\gamma_{a}|a_{m}|^{2}-ig_{r}b_{m}^{*}a_{m}-i(g_{l}-\xi_{m})b_{m-1}^{*}a_{m}\\
&\quad\,+\gamma_{a}|a_{m}|^{2}+ig_{r}b_{m}a_{m}^{*}+i(g_{l}-\xi_{m})b_{m-1}a_{m}^{*}\\
&\quad\,-ig_{r}a_{m}^{*}b_{m}-i(g_{l}-\xi_{m+1})a_{m+1}^{*}b_{m}\\
&\quad\,+ig_{r}a_{m}b_{m}^{*}+i(g_{l}-\xi_{m+1})a_{m+1}b_{m}^{*}\Big]\\
&=-\sum_{m}2\gamma_{a}|a_{m}|^{2}.
\end{split}
\label{A2}
\end{equation}
Similarly, one can verify that models B, C, and D also respect the norm evolution in Eq.~(\ref{A1}) due to the symmetric couplings. 

On the other hand, the RL model could also be extended to include asymmetric couplings. As an example, one could consider a model described by the dynamic equations
\begin{equation}
\begin{split}
\frac{\partial a_{m}}{\partial t}  &  =-\gamma_{a}a_{m}+i(\mu-\chi_{a,m})b_{m}+i(\nu-\chi_{a,m})b_{m-1},\\
\frac{\partial b_{m}}{\partial t}  &  =i(\mu-\chi_{b,m})a_{m}+i(\nu-\chi_{b,m})a_{m+1}
\end{split}
\label{A3}
\end{equation} 
with $\chi_{\alpha,m}=U|\alpha_{m}|^{2}$ ($\alpha=a,\,b$), which might be viewed as a simple dimerization of the Ablowitz-Ladik equation~\cite{NDDE,ALpra,renormAL}. For such a model, however, the evolution of the norm cannot be simply described by a sum over local decays from each lossy site and the definition of the mean displacement given in Eq.~(\ref{eq3}) cannot be applied.

\section{Coherent vs incoherent transport}\label{appb}

It has been shown that the mean displacement $\langle\Delta m\rangle$ displays continuous dependence on the relative values of $\nu$ and $\mu$ in the case of incoherent hopping and in the absence of nonlinearities~\cite{RLmodel}. Moreover, similar incoherent transport behavior has been obtained with on-site nonlinearities due to the interaction-induced decoherence effect~\cite{Rapedius}. The underlying physics of such results is quite useful for understanding the positive and negative fractional values of $\langle\Delta m\rangle$ in Fig.~\ref{fig2}.

For incoherent hopping, the original RL model exhibits continuous variation of $\langle\Delta m\rangle$ from $0$ to $1$ and thus can also take fractional values. This arises from the competition between the probabilities of the particle hopping from the initial site to all other sites. The particle initially located at site $|m=0,b\rangle$ can hop to either $|m=0,a\rangle$ or $|m=1,a\rangle$, corresponding to $\langle\Delta m\rangle=0$ and $\langle\Delta m\rangle=1$, respectively. However, the contributions of all subsequent hoppings should cancel out in this case (the hopping probabilities from $|m=0,b\rangle$ to $|m=-1,b\rangle$ and to $|m=+1,b\rangle$ are the same). In view of this, the value of $\langle\Delta m\rangle$ is determined by the probability of hopping to $A_{1}$ initially, which is given by $\langle\Delta m\rangle=\nu^{2}/(\nu^{2}+\mu^{2})$~\cite{RLmodel,Rapedius}. 

For models $A$ and $B$ here considered, the fractional values of $\langle\Delta m\rangle$ can be understood in a similar way. However, the most significant difference compared with the linear model above is that the contributions of long-range hopping processes no longer exactly cancel out due to the $m$-dependent effective couplings, i.e., the effective couplings on the left and right sides of the initial site are no longer symmetric. In particular, also negative fractional values of $\langle\Delta m\rangle$ are allowed, if the left-hand effective couplings are stronger than the right-hand ones.   

\section{Nonlinear couplings with only neutral-site contributions}\label{appc}

\begin{figure}[ptb]
\includegraphics[width=8.5cm]{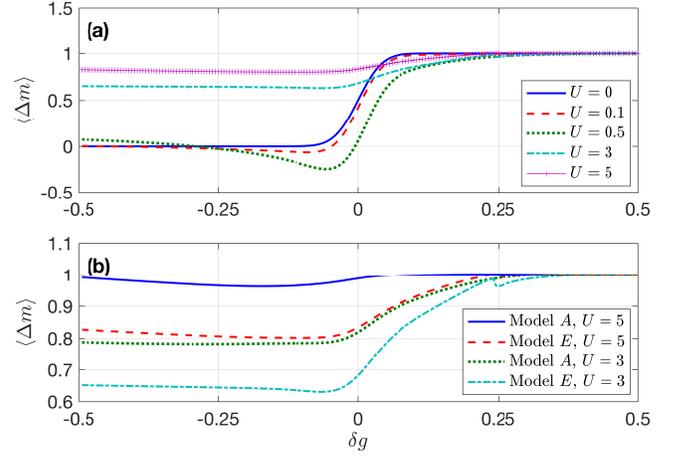} 
\caption{(Color online) (a) Mean displacement $\langle\Delta m\rangle$ of the particle versus coupling imbalance $\delta g$ for model $E$ with different values of $U$. (b) Comparison of $\langle\Delta m\rangle$ versus $\delta g$ for models $A$ and $E$ with relatively strong nonlinearities. The initial state is assumed as $|\psi(t=0)\rangle=|m=0,b\rangle$. Other parameters are $\mu=0.5-\delta g$, $\nu=0.5+\delta g$ and $\gamma_{a}=2$.}
\label{figa}
\end{figure}

In this appendix, we consider an extended RL model (model $E$) where the nonlinear intercell couplings only depend on the neutral-site field amplitudes, described by the dynamic equations  
\begin{equation}
\begin{split}
\frac{\partial a_{m}}{\partial t}  &  =-\gamma_{a}a_{m}+i\mu b_{m}+i(\nu-\eta_{m-1})b_{m-1},\\
\frac{\partial b_{m}}{\partial t}  &  =i\mu a_{m}+i(\nu-\eta_{m})a_{m+1}
\end{split}
\label{C1}
\end{equation} 
with $\eta_{m}=U|b_{m}|^{2}$ in this case.

Due to the fact that the particle is initially located at a neutral site with $|\psi(t=0)\rangle=|m=0,b\rangle$, the nearest-neighbor intercell coupling can still be markedly modified at the initial time by the local field amplitude $b_{0}$, similar to the case of model $A$. In view of this, Fig.~\figpanel{figa}{a} shows qualitatively the same behaviors as those in Fig.~\figpanel{fig2}{a}: as $U$ increases, the curve of $\langle\Delta m\rangle$ versus $\delta g$ tends to be flat, with the values approaching unity over the whole parametric range. In other words, the ``trivial-nontrivial'' transition in model $A$ can still be achieved even if the nonlinear intercell couplings only depend on the neutral-site field amplitudes. As shown in Fig.~\figpanel{figa}{b}, however, the effective intercell couplings are a bit weaker due to the absence of the contributions of the lossy sites, such that the values of $\langle\Delta m\rangle$ for model $E$ are smaller than those for model $A$, especially in the region of $\delta g<0$. To achieve similar modification effects, stronger nonlinearities are required in this case (see for instance the red dashed and green dotted lines).

\end{document}